\documentclass[twocolumn,showpacs,preprintnumbers,amsmath,amssymb,superscriptaddress]{revtex4-2}
\usepackage{graphicx}
\usepackage{dcolumn}
\usepackage{bm}
\usepackage{CJK, CJKnumb}
\usepackage{color}
\begin{document}
%\begin{CJK*}{GBK}{song}

\title{Band bending and zero-conductance resonances controlled by edge electric fields in zigzag silicene nanoribbons}
\author{{ Wei-Tao Lu$^{1,2}$, Qing-Feng Sun$^{3,4,5}$, Hong-Yu Tian$^{2}$, Ben-Hu Zhou$^{6}$, Hong-Mei Liu$^{2}$ }\\
\normalsize{${^1}^{}$\emph{Department of Physics, Yantai University, Yantai 264005, China}} \\
\normalsize{${^2}^{}$\emph{School of Physics and Electronic Engineering, Linyi University, Linyi 276005, China}} \\
\normalsize{${^3}^{}$\emph{International Center for Quantum Materials, School of Physics, Peking University, Beijing 100871, China}} \\
\normalsize{${^4}^{}$\emph{Collaborative Innovation Center of Quantum Matter, Beijing 100871, China }} \\
\normalsize{${^5}^{}$\emph{Beijing Academy of Quantum Information Sciences, West Bld. $\#$3, No. 10 Xibeiwang East Road, Haidian District, Beijing 100193, China }} \\
\normalsize{${^6}^{}$\emph{Department of Physics, Shaoyang University, Shaoyang 422001, China }} \\}

\begin{abstract}
We study the band structure and transport property of a zigzag silicene nanoribbon when the electric fields are applied to the edges. It is found that a band bending could be induced and controlled by the antisymmetric edge fields, which can be understood based on the wave functions of the edge states. The highest valence band and the lowest conduction band coexist in the band bending region. With the narrowing of edge potentials, the bending increases gradually. When the edge fields become symmetric, an asymmetric band gap at the Dirac points can be obtained due to the intrinsic spin-orbit interaction, suggesting a valley polarized quantum spin Hall state. The gap could reach a maximum value rapidly and then decrease slowly as the electric fields increase. Due to the combining effect of the band bending, band selective rule, and resonant states, many zero-conductance resonances and resonance peaks appear in different regions, which could be described by the Fano resonance effect. Furthermore, the band bending and zero-conductance resonances are robust against the Hubbard interaction. The Hubbard interaction could work as a spin-dependent edge field, together with the edge electric fields, leading to a spin-dependent band gap and various quantum phases such as metal and half-metal.
\end{abstract}
\maketitle

\section{Introduction}

Recently, silicene has attracted much attention both theoretically and experimentally \cite{Voon,Houssa}, which could be potentially integrated with the well established silicon technology. Contrary to graphene, silicene has a strong intrinsic spin-orbit interaction (SOI), which could open a gap of approximately $1.55meV$ at Dirac points \cite{Liu,Drummond}. The buckled structure of silicene allows us to control the band gap by an external electric field, offering great advantages over the gapless graphene \cite{Drummond,Ezawa}. Therefore, many novel properties of silicene are demonstrated, including quantum spin Hall effect and the topological phase transition by applying electric field \cite{Liu,Ezawa}.

Numerous literatures have researched the band structures and conductance property in graphene nanoribbons (GNRs) \cite{Klein, Fujita, White, Louie, Louie2, Louie3, Brey, Nakabayashi, Hod, Martins, Long, Sun, Rakyta, Duan, Kim, LiLu, Rycerz, Rossier, Mucciolo, Gunlycke, Niu, Cheng} and silicene nanoribbons (SiNRs) \cite{Xu, Kang, Rachel, An, Tsai, Nunez, Rzeszotarski, Lu, Pan, Li, Zhang, Zberecki, Nagaosa, Shakouri, Woods}. Fujita found that GNRs display striking contrast in the electronic states depending on the edge shape \cite{Fujita}. Louie et al. predicted the half-metallicity in nanometre-scale GNRs by using first-principles calculations, opening a new path to explore spintronics \cite{Louie}. Because of the staggered sublattice potential on the hexagonal lattice due to edge magnetization, an energy gap could be opened for zigzag GNRs \cite{Louie2} and the spin-polarized valley helical edge states could appear inside the bulk gap \cite{Niu}. If the Rashba SOI is stronger than the intrinsic SOI, the low-energy bands would undergo trigonal-warping deformation at the Dirac points \cite{Rakyta}. In zigzag GNRs, the scattering processes obey a selection rule for the band indices and so a barrier potential can play the role of the band-selective filter \cite{Nakabayashi}. Due to the coupling between the conducting subbands around the Fermi level, GNRs show distinctly different transport behaviors, depending on whether they are mirror symmetric with respect to the midplane between two edges \cite{Duan}. Long et al. found that the conductance through graphene $p-n$ junctions could be strongly enhanced and exhibit a plateau structure at a suitable range of disorders \cite{Long}. Magnetoresistance effect \cite{Kim,Rossier} and valley filtering \cite{Rycerz,Gunlycke,Cheng} in GNRs device have been demonstrated. Furthermore, magnetoresistance effect \cite{Xu,Kang,Rachel}, spin-polarized currents \cite{An,Tsai,Nunez,Rzeszotarski,Lu}, valley-resolved transport \cite{Pan,Li}, electron delocalization \cite{Zhang}, and thermoelectric effects \cite{Zberecki} in SiNRs have also been studied theoretically. In particular, Rachel and Ezawa found that the quantum spin Hall effect without edge states could be realized by manipulating various perturbations at the edges of SiNRs, which can be used for giant magnetoresistance and spin filter \cite{Rachel}. The silicene-based spin-filter and Y-shaped spin/valley separator are proposed in field-gated SiNRs by first-principles calculations \cite{Tsai}.

The edge potential is an effective method of engineering the electronic structure in GNRs \cite{White2, White3, Yao, Zhang2, Bhowmick, Apel, Chiu, Lee, Kumada}. The electronic structure close to the Fermi level greatly depends on the choice of terminating atom or group at the edges of GNRs, leading to the particular device applications by chemically modifying the edges \cite{White2, White3}. The edge band dispersion of zigzag GNRs can be controlled by potentials applied on the boundary with unit cell length scale, and the gapless edge states with valley-dependent velocity are found \cite {Yao}. When the applied edge potential is antisymmetric, the GNRs energy spectrum could open up a gap \cite{Apel}. A recent study investigated topological confinement effects of the edge potentials resulting from electron-electron interactions on gapless edge states in various magnetic phases of bilayer zigzag GNRs \cite{Lee}. Although the armchair GNRs have no edge states, it is possible to generate and tune the edge states due to pseudospin-flipped scattering induced by the edge potential \cite{Chiu}. Experimental investigation indicated that owing to the sharp edge potential and the linear band structure, resonant edge magnetoplasmons dissipation in graphene can be lower than that in GaAs systems \cite{Kumada}.

In this work, we propose a different kind of edge potential in zigzag SiNRs (ZSiNRs) produced by the electric fields $E_{z1,z2}$ applied to the edges of the ribbon, as shown in Fig. 1(a), which is different from previous works where the electric fields are usually applied to the whole ribbon. Due to the buckled structure of silicene, the electric fields could generate a staggered sublattice potential $e \ell E_{z1,z2}$ with $2 \ell \approx 0.46 {\AA}$ the vertical separation of $A$ and $B$ sites of the two sublattices, which critically distinguishes silicene from graphene. The edge potentials could greatly affect the edge states and the band structure. We aim to study the energy band and the conductance in ZSiNRs controlled by the edge potentials $e \ell E_{z1,z2}$ as well as discuss the effect of the intrinsic SOI. It is interesting that the band can be bent and gapped by the local manipulations of edge potentials, depending on the symmetry of edge potentials $e \ell E_{z1,z2}$. The conductance presents abundant transport behaviors, such as zero-conductance resonances and resonance peaks. The corresponding density of states (DOS) and local density of states (LDOS) are discussed. Contrary to previous literatures \cite{Rakyta, Faria}, the band bending and zero-conductance resonances are electrically tunable in this work. Taking the exchange field into account, the zero-conductance resonances could give rise to a perfect spin polarization. In addition, we also study the effect of Hubbard interaction and discuss the spin-dependent band gap and the half-metallicity of ZSiNRs.

The paper is organized as follows. In Sec. II we introduce the effective tight-binding model and the nonequilibrium Green's function (NEGF) method. The numerical results and the discussions are presented in Sec. III. We conclude with a summary in Sec. IV.

\section{Theoretical Formulation}

The low-energy electronic properties in the silicene honeycomb lattice can be described very well by the tight-binding model \cite{Ezawa,Liu2}. Considering the electric fields, the Hamiltonian reads
\begin{eqnarray}
&&  H_{TB} = -t \sum_{\langle i,j \rangle, \alpha} c_{i \alpha}^{\dag} c_{j \alpha} + i \frac{\lambda_{SO}}{3\sqrt{3}}
    \sum_{\langle\langle i,j \rangle\rangle, \alpha, \beta} v_{ij} c_{i \alpha}^{\dag} (\sigma_z)_{\alpha\beta} c_{j \beta} \nonumber\\
&&  + \sum_{i,\alpha} e \ell E_{z1,z2} \xi_i c_{i \alpha}^{\dag} c_{i \alpha}.
\end{eqnarray}
The first term describes the nearest-neighbor hopping with the transfer energy $t=1.6 eV$, and $c_{i \alpha}^{\dag}$ ($c_{i \alpha}$) is the electronic creation (annihilation) operator with spin $\alpha$ at site $i$. The second term is intrinsic SOI with $\lambda_{SO} = 3.9meV=0.0024t$, which involve spin dependent next nearest-neighbour hopping. $v_{ij} = +1 (-1)$ if the next nearest-neighboring hopping is anticlockwise (clockwise) with respect to the positive $z$ axis. $\sigma_z$ is the Pauli matrix associated with spin degree of freedom. The third term describes the staggered sublattice potential $U_{E1,E2}=e \ell E_{z1,z2}$ arising from the electric fields $E_{z1,z2}$ perpendicular to the silicene sheet, as shown in Fig.1(a). The potentials $U_{E1,E2}$ are applied along two edges of ZSiNRs in the center region, i.e., edge potential, the size of which is described by the width $W$ and length $N_x$. Fig. 1(b) shows that the band structure of the pristine ZSiNRs with width $N_y=48$ and potential $U_{E1,E2}=0$, which exhibits a metallic behavior, similar to graphene. The lowest bands at the Fermi level are not absolutely flat due to the intrinsic SOI, associating with the helical edge states.

\begin{figure}
\includegraphics[width=8.0cm,height=6.0cm]{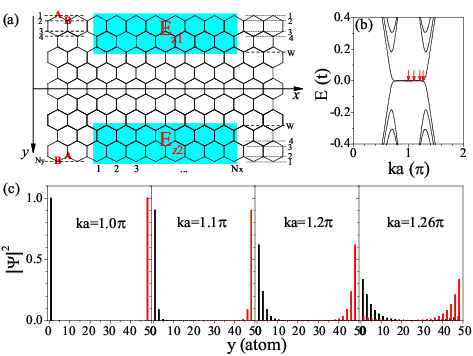}
\caption{ (a) The schematic diagram of a silicene device with zigzag ribbon, on the two sides of which two electric fields $E_{z1,z2}$ are placed. $N_x$ is the number of unit cells for center region in the $x$ direction. The unit cell includes $N_y$ silicon atoms which defines the ribbon width. $N_x$ and $W$ describe the region of the electric fields. The band structure of a pristine zigzag ribbon for $N_y=48$ is shown in (b). (c) The probability density for the wave functions $|\Psi|^2$ of the edge states labeled by the arrows in (b), where the black and red lines are the states at the upper and lower boundaries of the ribbon, respectively.}
\end{figure}

Based on the tight-binding model and Bloch's theorem, the band structure of an infinite ZSiNRs with edge potentials can be calculated. The $k$-dependent Hamiltonian of the system can be written as
\begin{eqnarray}
H_k = H_{00} + H_{01} e^{ika} + H_{-10} e^{-ika},
\end{eqnarray}
where $H_{00}$ is a unit cell Hamiltonian matrix of one chain, $H_{01}$ (or $H_{-10}$) is the coupling matrix with the right-hand (or left-hand) adjacent cell, and $a$ is the lattice constant. In addition, the two-terminal conductance $G$ for an electron with energy $E$ through the silicene ribbon can be calculated by the NEGF method and the Landauer-B\"{u}ttiker formula as \cite{Datta,Long,Sun}
\begin{eqnarray}
G(E) = \frac{e^2}{h} Tr [\Gamma_L(E) G^r(E) \Gamma_R(E) G^a(E)],
\end{eqnarray}
where $\Gamma_{L,R}(E) = i [\Sigma_{L,R}(E) - \Sigma^{\dag}_{L,R}(E)]$ is the linewidth function and $G^r(E) = [G^a(E)]^{\dag} = 1 / (E - H_c - \Sigma_L - \Sigma_R)$ is the retarded Green function with the Hamiltonian in the center region $H_c$. $\Sigma_{L,R}$ is the selfenergy caused by the coupling between the center and lead regions. The local density of states (LDOS) and density of states (DOS) can also be given by the NEGF method \cite{Wang}:
\begin{eqnarray}
&& LDOS = - \frac{1}{\pi} Im G^r(E)    \nonumber\\
&&  = \frac{1}{2\pi} [G^r(E)(\Gamma_L(E)+\Gamma_R(E))G^a(E)]
\end{eqnarray}
and
\begin{eqnarray}
DOS = \frac{1}{2\pi} Tr[G^r(E)(\Gamma_L(E)+\Gamma_R(E))G^a(E)].
\end{eqnarray}

\section{Results and Discussions}

In this section, we study the effect of the edge electric fields and intrinsic SOI on the band bending, band gap, resonant conductance, and spin polarization for a zigzag silicene nanoribbon by calculating Eqs. (2)-(5). Note that in figures 2-6, we mainly discuss the results of spin up electrons. Because of time-reversal symmetry, the band structures of spin up and spin down electrons satisfy $E_{\uparrow}(k)=E_{\downarrow}(-k)$, and so conductance is spin independence. However, when the Hubbard interaction is considered, the system presents a spin-dependent band gap and phase transition controlled by the edge electric field.

\subsection{Band bending and band gap}

\begin{figure}
\includegraphics[width=8.0cm,height=7.0cm]{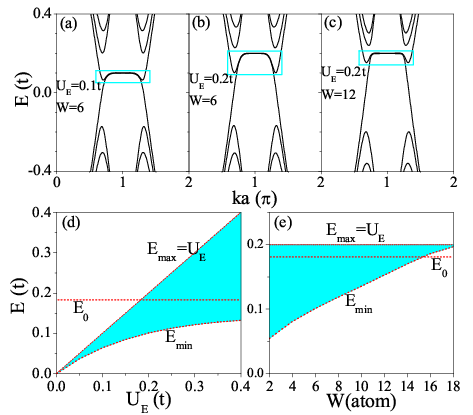}
\caption{ (a)-(c) The band structure of ZSiNRs for different values of the sublattice potential $U_E$ and the width $W$. The position and the width of the band bending region as a function of (d) $U_E$ with $W=6$ and (e) $W$ with $U_E=0.2t$. The horizontal line denotes the position of the bottom for the next lowest conduction band $E_0=0.183t$ in the pristine zigzag ribbon. Here, the ribbon width $N_y=48$. }
\end{figure}

Firstly, we discuss the band structure of an infinitely long nanoribbon in the present of the edge electric fields where the length $N_x$ tends to infinite, as shown in Figs. 2 and 3. Note that the edge fields applied to the two edges are antisymmetric in this case, that is $\vec{E}_{z1}=-\vec{E}_{z2}=E_z\vec{z}$, but their corresponding sublattice potentials $U_{E1,E2} = U_E$ at the edges are symmetric about the $x$ axis and independent of spin degree of freedom. Figs. 2(a)-2(c) exhibit the energy band for different values of the edge potentials $U_E$ and the width $W$. One can clearly see that the lowest conduction band presents a bend behaviour due to the appearance of the sublattice potential $U_E$. Importantly, the maximum energy of the highest valence band becomes larger than the minimal energy of the lowest conduction band. Therefore, there exist an especial energy region where both the highest valence band and the lowest conduction band coexist and overlap, which plays a key role to the resonant conductance as shown in the following. Comparison between Figs. 2(a) and 2(b) indicates that as the potential $U_E$ increases, the band bending region increases and shifts up. With the increase of width $W$, the band bending is decreased (see Figs. 2(b) and 2(c)). The further increase of $U_E$ could induce the bend of the next lowest conduction band and other higher band. It should be noted that the valence band could also be bent when the edge potentials are negative. Figs. 2(d) and 2(e) present the band bending region as a function of $U_E$ and $W$, respectively. The boundary of bending region is denoted by $E_{min}$ and $E_{max}$ with $E_{max}=U_E$. $E_0$ is the position of the bottom for the next lowest conduction band in the pristine zigzag ribbon (or the two leads) which can be controlled by the ribbon width $N_y$. It is clearly seen that the bending region is broadened (or narrowed) gradually by $U_E$ (or $W$). When $W$ increases to $N_y/2$, the two edge potentials would be combined into one potential which covers the whole area of the nanoribbon, and so the band bending disappears. Therefore, the position and the width of band bending can be controlled by the edge electric fields.

The physical mechanism of band bending can be understood in the light of wave functions of the edge states. We assume the sublattices on the upper and lower edges of the ZSiNRs are A and B, respectively, as shown in Fig. 1(a). Fig. 1(c) plots the probability density profile of edge-state wave functions $|\Psi|^2$ as a function of the atom positions along the width of the pristine silicene sheet, and the corresponding edge states are labeled in Fig. 1(b). It can be seen that at symmetric point $ka=1.0\pi$, the states are only localized at the edges. However, with the increase of $ka$, the states are extended to the whole ribbon. Importantly, the state on the upper edge is mainly localized at A sublattice, while the state on the lower edge is mainly localized at B sublattice. For the antisymmetric edge electric fields, the potential $U_E$ at A sublattice on the upper edge is same as that at B sublattice on the lower edge. When the potential $U_E$ with a narrower width is applied to the edges, it would greatly affect the states near $ka=1.0\pi$ but have little effect on the states far away from $ka=1.0\pi$. Thus, the lowest conduction band and the highest valence band near $ka=1.0\pi$ shift up linearly with $U_E$, while the bands remote from $ka=1.0\pi$ shift slowly. As a consequence, the band bending appears. As the width $W$ broadens, the states and the bands far away from $ka=1.0\pi$ are influenced deeply as well, and so the band bending becomes narrow, consistent with the results in Fig. 2.

\begin{figure}
\includegraphics[width=8.0cm,height=7.0cm]{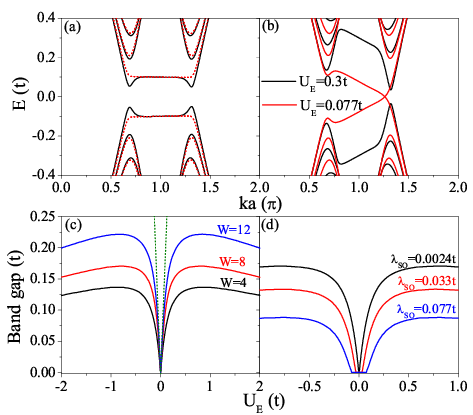}
\caption{ The band structure of ZSiNRs for (a) $U_E=0.1t$, $\lambda_{SO} = 0.0024t$ and (b) $U_E=0.3t$, $\lambda_{SO} = 0.077t$. (c)(d) Band gap versus $U_E$. The dashed curves in (a) and (c) denote the band structure and band gap for $W=N_y/2$, respectively. Here, $N_y=48$ and $W=8$ unless otherwise noted. }
\end{figure}

The symmetry of the edge electric fields greatly affects the band structure of ZSiNRs. When the edge electric fields become symmetric ($\vec{E}_{z1}=\vec{E}_{z2}=E_z\vec{z}$), the corresponding sublattice potentials are antisymmetric about the $x$ axis. The potential $U_E$ at A sublattice on the upper edge is positive while the potential at B sublattice on the lower edge is negative (see Table I). Thus, the staggered sublattice potentials between A and B sublattices on the edges would separate the conduction band and valence band, opening a band gap. Fig. 3(a) displays the band structure of ZSiNRs when the potential $U_E=0.1t$ and $W=8$. We may find that a spectral gap is induced, and the system converts metallic to insulating behavior. For comparison, when the electric field is applied to the whole area of the nanoribbon, i.e., $W=N_y/2$,  the band structure and band gap are presented by the dashed curves in Figs. 3(a) and 3(c), respectively. Obviously, the energy band is bent near the valleys when the edge potentials are applied (see Fig. 3(a)), which can be understood by the edge states in Fig. 1(c). Furthermore, in contrast to GNRs \cite{Apel}, the gap width of ZSiNRs at the two Dirac points is not consistent, and the two valleys is asymmetrical due to the intrinsic SOI. For stanene with a larger SOI $\lambda_{SO} = 0.077t$, the asymmetry of the valleys are more prominent, as shown in Fig. 3(b). When $U_E=\lambda_{SO}$, the gap at one valley is closed and a valley polarized quantum spin Hall state could be formed in ZSiNRs. This feature could be used to realize a valley polarization, where the conductance are only contributed by electrons from one valley. Fig. 3(c) shows the band gap as a function of $U_E$ for different value of $W$ when $N_y=48$. With the increase of $U_E$, the band gap increases rapidly and reaches the maximum value at a special value of $U_E$. Then the gap decreases slowly with $U_E$. Interestingly, the maximum of band gap is dependent of the width $W$. Note that this property is different from the one when the electric field is applied to the whole ribbon, where the gap is increased linearly with $U_E$ and equal to $2 U_E$ (see the dashed curve in Fig. 3(c)). Fig. 3(d) discusses the effect of the intrinsic SOI on the band gap with $\lambda_{SO}=0.0024t$, $0.033t$, and $0.077t$ in silicene, germanene, and stanene. We can find that when $U_E<\lambda_{SO}$, the gap cannot be opened up. When $U_E>\lambda_{SO}$, the gap is opened up and increases gradually. Due to the SOI, the band structure is no longer symmetric, and so the indirect band gap is reduced, which is proportional to $1/\lambda_{SO}$.

\subsection{Zero-conductance resonances and spin filter}

\begin{figure}
\includegraphics[width=8.0cm,height=8.0cm]{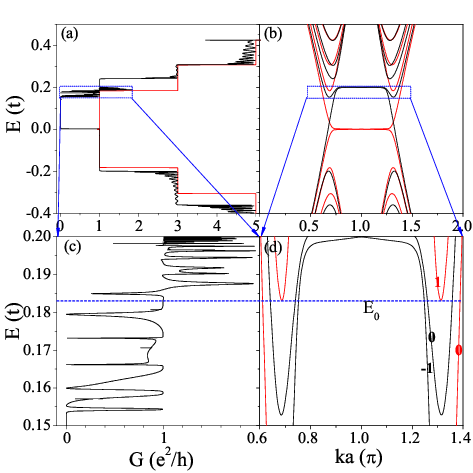}
\caption{ (a) The conductance $G$ versus Fermi energy through the system labelled by black curves, and the red curves are the conductance of the pristine zigzag ribbon. (b) The band structures of the center region (black curves) and the two leads (red curves). (c) and (d) are the partial enlargements of (a) and (b), respectively. The subbands are indexed in (d). The parameters are set as $U_E=0.2t$, $W=12$, $N_x=100$, and $N_y=48$. }
\end{figure}

The band bending would further affect the electron transport through the silicene nanoribbon. Next, we discuss the conductance through a finite ZSiNRs and the effect of band bending when the edge electric fields are applied to the edges antisymmetrically. The width of ZSiNRs and the potential are fixed as $N_y=48$ and $U_E=0.2t$ in the following for convenience. Fig. 4 presents (a)(c) the conductance and (b)(d) the corresponding band structure when the width and the length of edge electric fields are $W=12$ and $N_x=100$. In order to contrast, the conductances for the considered system and the pristine ZSiNRs are labeled by the black and red curves in Fig. 4(a), respectively. The black and red curves in Fig. 4(b) represent the energy bands for the considered system in the center region and the two leads (or the pristine ZSiNRs), respectively. As expected, the quantized conductance in pristine ZSiNRs is obtained, which exhibits a symmetric plateau structure with plateau values at $1$, $3$, $5$, $\cdots$ in the unit $e^2/h$. The conductance plateau is proportional to the number of conducting channels at the Fermi energy, consistent with the number of subbands at the Fermi energy. Oppositely, the conductance in the considered system presents a resonance behavior which is no longer symmetric. The features of resonant conductance for electrons and holes are different in the multi-channel region. The conductance is zero in the region $0<E<E_{min}$ due to the selection rule for the band indices. In the following, we mainly focus on the resonant conductance in the region $E_{min}<E<E_{max}$ where the band bending occurs, as shown in Fig. 4(c). Remarkably, it can be found that many resonance dips with conductance $G=0$ appear in the region $E_{min}<E<E_0$. In the region $E_0<E<E_{max}$, the resonance dips and the resonance peaks occur simultaneously, and the peak value trend to $2e^2/h$. This phenomenon can be understood based on the band bending, band selective rule, and resonant states. The corresponding energy band of the center region and two leads is shown in Fig. 4(d), and the band index is specified. The electrons in the even (odd) bands can be scattered only into the even (odd) bands due to the conservation of the parity of wave functions \cite{Nakabayashi,Chen}. On the other hand, the highest valence band and the lowest conduction band are coexistent in the band bending region. In the center region, the bands $0$ and $-1$ are coexistent at $E_{min}<E<E_{max}$ (see Fig. 4(d)). In the left and right leads, only the band $0$ exists at $E_{min}<E<E_0$, while the bands $0$ and $1$ coexist at $E_0<E<E_{max}$. Therefore, when the electron with energy $E \in [E_{min}, E_0]$ passes through the system, the band $0$ in the leads is scattered into the band $0$ in the center region, so that the conductance $G$ is usually $e^2/h$ and the plateau arises from nonresonant electron traveling. However, when the incident energy $E$ consists with energy of resonant modes in the band $-1$ induced by the edge potentials, the conductance would drop to zero due to the opposite parity between the band $0$ in the lead and the band $-1$ in the center region, resulting in zero-conductance resonance. On the contrary, for incident energy $E \in [E_0, E_{max}]$, the electron in the band $1$ in the lead is consistent with the band $-1$ in the center region, and so the same parity leads to the emergence of resonance peak at resonant energy in the conductance.

\begin{figure}
\includegraphics[width=8.0cm,height=7.0cm]{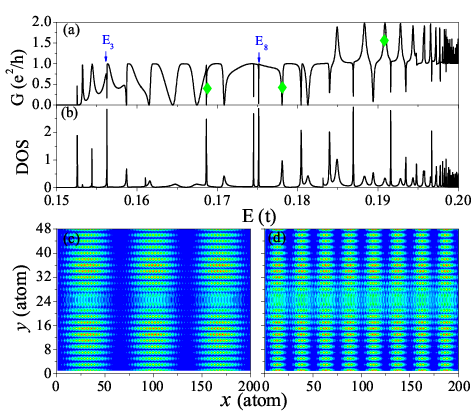}
\caption{ (a) The conductance and (b) DOS versus Fermi energy at $N_x=200$. LDOS in the center region at (c) $E=E_3=0.15625t$ and (d) $E=E_8=0.17515t$. Other parameters are the same as these in Fig.4.}
\end{figure}

Fig. 5 presents (a) the conductance $G$ and (b) the corresponding DOS at $N_x=200$. Obviously, the resonance dips and resonance peaks of conductance correspond exactly to the peaks of DOS. In addition, the comparison between Figs. 5(a) and 4(c) manifests that as the length of the center region increases, the number of conductance resonance gradually increases, because more resonant modes appear in the center region. When $E=E_3$ and $E_8$, the LDOS in the center region is painted in Figs. 5(c) and 5(d), respectively, corresponding to the certain resonance dips in Fig. 5(a). Clearly, the standing waves are formed at the zigzag edges in the center region. When Fermi energy is outside the band bending range, i.e., $E<E_{min}$ or $E>E_{max}$, there is no bound state, and so the conductance is $0$ or $e^2/h$ (see Fig. 4(a)). This result proves that the bound state induced by edge potential offers a resonant channel for conductance.

\begin{figure}
\includegraphics[width=8.0cm,height=6.0cm]{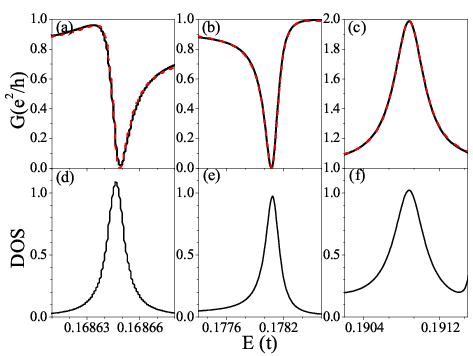}
\caption{ (a)-(c) The Fano resonance peaks marked in Fig. 5(a) obtained by the NEGF method (black curves) and fitted by the normalized Fano curve (red dashed curves). (d)-(f) DOS corresponds to Fano resonance in (a)-(c). }
\end{figure}

In fact, the above resonance phenomenon of conductance can be described by Fano resonance effect \cite{Miroshnichenko}. Fano resonance is a kind of asymmetric resonance which arises from the quantum interference effects between the discrete and continuous states. Recently, Fano resonance of conductance controlled by the magnetic fields in hexagonal zigzag graphene rings \cite{Faria} and bilayer phosphorene nanoring \cite{Zhang3} has been researched. The formula for the shape of the Fano resonance profile can be expressed as
\begin{eqnarray}
T(E) = \frac{T_0}{1+q^2} \frac{(\epsilon+q)^2}{1+\epsilon^2},
\end{eqnarray}
with $\epsilon=(E-E_F)/\Gamma$ \cite{Miroshnichenko}. $E_F$ and $\Gamma$ determine the position and width of the Fano curve, respectively. $T_0$ is the maximum value of the peak and $q$ is a phenomenological shape parameter. Figs. 6(a)-6(c) show the Fano resonance curves labeled by rhombuses in Fig. 5(a) and calculated by equation 6, which match each other perfectly. In particular, the conductance in Fig. 6(a) clearly exhibits a sharp dip is followed by a sharp peak, which is the typical characteristic of asymmetric Fano resonance. For proper Fermi energy in the band bending region, the edge electric fields could induce the bound states at the edges (see Fig. 5). Meanwhile, the states far away from $ka=1.0\pi$ are always continuous states which are distributed throughout the ribbon (see Fig. 1(c)). The bound states at the edges would interfere with the continuum states in the whole ribbon, leading to the occurrence of Fano resonance in conductance. The corresponding DOS of Fano resonance is shown in Figs. 6(d)-6(f). One can see that the positions of the DOS peaks match very well with the Fano resonances.

\begin{figure}
\includegraphics[width=8.0cm,height=5.0cm]{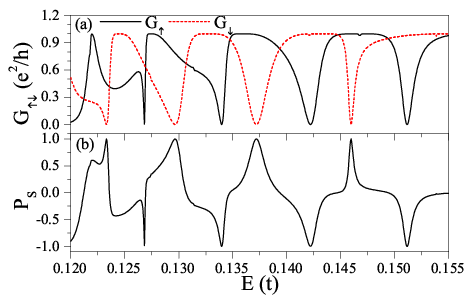}
\caption{ (a) The spin conductance $G_{\uparrow,\downarrow}$ and (b) spin polarization $P_S$ versus Fermi energy, at $W=8$, $N_x=100$, and $M=0.004t$. }
\end{figure}

As an application of the zero-conductance Fano resonance, the system could work as a controllable spin filter by introducing an exchange field $M$. In a real experimental condition, the exchange field $M$ may be quite small. For instance, by growing the graphene on a ferromagnetic insulator such as EuO, its magnetization $M$ is only about $5 meV$ \cite{Haugen}. Thus, in this system, we consider a small exchange field with $M=0.004t=6.4meV$, which is also applied on the edges of ZSiNRs, the same as the potential $U_E$. The corresponding Hamiltonian is
\begin{eqnarray}
H = H_{TB} + \sum_{i, \alpha} M_i c_{i \alpha}^{\dag} \sigma_z c_{i \alpha}.
\end{eqnarray}
The spin-dependent conductance as a function of incident energy is shown in Fig. 7(a). It is evident that both spin up and spin down electrons present the zero-conductance resonance in the region $E_{min}<E<E_0$. The characteristic of spin transport can be obviously seen from the spin polarization defined by $P_S=(G_{\uparrow}-G_{\downarrow})/(G_{\uparrow}+G_{\downarrow})$, as shown in Fig. 7(b), illuminating that the $100\%$ spin polarization can be reached within certain energies. The polarization changes sign when the energy varies between the values that give the resonances for both spin orientations. Therefore, it is possible to realize a spin polarization switch by the application of the zero-conductance resonance and weak exchange splitting.

\subsection{The effect of Hubbard interaction}

In the preceding subsections we have neglected the Hubbard interaction. In fact, this interaction can have a significant
impact not only on the band structure but also on the transport properties. When spin is not considered, there is a doubly degenerate flat band for edge state at the Fermi level (see Fig. 1(b)), which will enhance the electron-electron interaction at the edges. As a consequence, the ZSiNRs are predicted to have an antiferromagnetic ground state with antiparallel spin orientation between the two edges \cite{Ni}. The band gap appears because of a staggered sublattice potential due to spin ordered states at the edges \cite{Louie2, Ni}. The electron-electron interaction is generally taken as the Hubbard interaction in the mean-field approximation. Including the Hubbard interaction, the Hamiltonian for ZSiNRs reads \cite{White3,White4}
\begin{eqnarray}
H_{Hubbard} = H_{TB} + U \sum_{i, \alpha} ( \langle n_{i -\alpha} \rangle - 1/2 ) n_{i \alpha}.
\end{eqnarray}
The last term describes Hubbard interaction and $U$ indicates the on-site Coulomb repulsion energy between the opposite spins. $U=1.4eV$ for ZSiNRs is parameterized from the results of spin-unrestricted first-principles calculation \cite{Wierzbicki}. $\langle n_{i \alpha} \rangle$ is the average local spin occupation on atom $i$ for spin $\alpha$ which is solved self-consistently. $n_{i \alpha}$ is the particle number operator.

\begin{table}
\caption{The sublattice potentials for spin-up and spin-down electrons at the A (B) sublattice on the upper (lower) edge of the ZSiNRs under the Hubbard term and the edge electric fields. The superposition $1$ ($2$) is the combine effect of the Hubbard term and the symmetric (antisymmetric) edge electric fields.}
\begin{center}
\begin{tabular}{|c|c|c|c|c|c|c|}\hline
Sublattice              &   A         &   A          &  B           &   B           \\ \hline
Spin index              &  spin up    &  spin down   &  spin up     &  spin down    \\ \hline
Hubbard term            &  $U_H$      &  $-U_H$      &  $-U_H$      &  $U_H$        \\ \hline
Field $E_{z1}=E_{z2}$   &  $U_E$      &  $U_E$       &  $-U_E$      &  $-U_E$       \\ \hline
Field $E_{z1}=-E_{z2}$  &  $U_E$      &  $U_E$       &  $U_E$       &  $U_E$        \\ \hline
Superposition $1$       &  $U_H+U_E$  &  $-U_H+U_E$  &  $-U_H-U_E$  &  $U_H-U_E$    \\ \hline
Superposition $2$       &  $U_H+U_E$  &  $-U_H+U_E$  &  $-U_H+U_E$  &  $U_H+U_E$    \\ \hline
\end{tabular}
\end{center}
\end{table}

Before proceeding with the calculation, we discuss the sublattice potentials induced by the Hubbard term and the edge electric fields. Analogy with the wave functions in Fig. 1(c), the spin occupation $\langle n_{i \alpha} \rangle$ implies that the spin-up (spin-down) electrons mainly accumulate on A (B) sublattice of the upper (lower) edge of ZSiNRs \cite{Louie, White3}. Thus, the Hubbard term could be regarded as a spin-dependent scalar field which is applied on A (B) sublattice of the upper (lower) edge, just like the edge electric fields. Table I displays the sublattice potential for spin up and spin down at the A (B) sublattice on the upper (lower) edge of the ZSiNRs when the Hubbard term and the edge electric fields are considered. We set two potential parameters $U_H=U ( \langle n_{1 \alpha} \rangle - 1/2 )$ and $U_E=e \ell E_z$ in order to describe the sublattice potential. For Hubbard term, the spin-dependent edge field could result in a spin-dependent sublattice potential on the edges, which has antisymmetric distribution with respect to the $x$ axis for both spins, where the potential is $U_H$ ($-U_H$) at A (B) sublattice on the upper (lower) edge for spin up, while the potential is $-U_H$ ($U_H$) at A (B) sublattice for spin down. The potentials between the two spins are symmetric about the $x$ axis. Thus, a spin-independent band gap is opened due to the staggered sublattice potential at the edges (see the magenta dash curves in Fig. 8(a)), consistent with the results by first-principles theory \cite{Ni}. The symmetric (antisymmetric) edge electric fields could induce a spin-independent sublattice potential which has antisymmetric (symmetric) distribution with respect to the $x$ axis (see Table I). Therefore, it is feasible to control the energy band of spin-polarized edge states by the Hubbard term and edge electric fields.

\begin{figure}
\includegraphics[width=8.0cm,height=6.0cm]{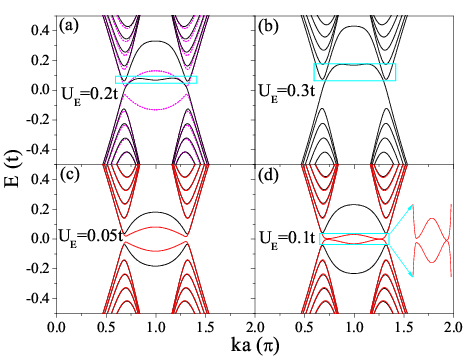}
\caption{ The band structure of ZSiNRs in the present of the (a)(b) antisymmetric and (c)(d) symmetric edge electric fields calculated by Hubbard model. Here, the ribbon width $N_y=64$ and field width $W=4$. The black curves and red curves in (c)(d) are for spin up and spin down, respectively. The magenta dash curves in (a) are for the pristine ZSiNRs when the Hubbard interaction is considered. }
\end{figure}

Fig. 8 shows the band structure of ZSiNRs calculated by Hubbard model. When the edge electric fields are antisymmetric, the combine effect of the two kinds of sublattice potentials generates a spin-dependent asymmetric edge potential (see superposition $2$ in Table I). This asymmetry between the A and B sublattices on the edges suggests the different sublattice potentials, leading to the separation of conduction and valence bands. When the potentials $U_H+U_E$ and $-U_H+U_E$ are positive at A and B sublattices, the conduction and valence bands of edge states can be shift up, leading to the bending of band. Meanwhile, the sublattice potentials between the two spins are symmetric around the $x$ axis, leading to the spin-independent band. Therefore, as shown in Figs. 8(a) and 8(b), a spin-independent band bending occurs where the highest valence band and the lowest conduction band coexist but separate, contrary to that observed in Fig. 2 by tight-binding model. The band bending is more remarkable as the electric field increases (see Fig. 8(b)). Furthermore, in the Hubbard model, the system undergoes a phase transition from semiconductor to metal due to the band bending. When the edge electric fields are symmetric, the total sublattice potentials for the two spins become antisymmetric about the $x$ axis, but the potentials between opposite spins are different (see superposition $1$ in Table I), which would generate a spin-dependent staggered sublattice potential. Thus, the spin-dependent band gaps are observed where the gap for spin up is broadened but the gap for spin down is narrowed (see Figs. 8(c) and 8(d)).

\begin{figure}
\includegraphics[width=8.0cm,height=4.0cm]{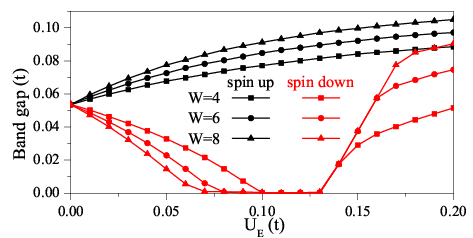}
\caption{ The band gap for spin up and spin down versus the symmetric edge electric fields calculated by Hubbard model with $N_y=64$. }
\end{figure}

Fig. 9 presents the band gap for spin up and spin down as a function of the symmetric edge electric fields for different width $W$. Under the edge fields, the band gaps of the spin up and spin down change differently. The band gap for spin down decrease to zero first and then increases up rapidly, while the gap for spin up increases monotonously with the electric field. In particular, when the electric field-induced potential $U_E=e \ell E_z$ approaches to the maximum of the Hubbard-induced potential $U_H \approx 0.13t$, the difference of the total potential between A and B sublattices for spin down would tend to zero, and so the band gap for spin down is closed while the gap for spin up is about $0.09t$. Thus, the ZSiNRs become half-metal. When $U_E>U_H$, the band gap is reopened up. With the increase of the width $W$, the effect of electric field becomes more and more prominent, and so the critical electric field for achieving half-metallicity decreases. Remarkably, the system undergoes a series of phase transition from semiconductor to half-metal to semiconductor. In particular, due to the intrinsic SOI of silicene, together with Hubbard term and edge electric fields, the spin degeneracy can be destroyed and the energy band is no longer symmetric near the two valleys. Therefore, at the critical value of the electric field, the band gap of a certain spin near a certain valley could be closed while other gaps remain. As a consequence, a spin and valley polarized quantum anomalous Hall state could be formed in ZSiNRs, as shown in Fig. 8(d).

\begin{figure}
\includegraphics[width=8.0cm,height=6.0cm]{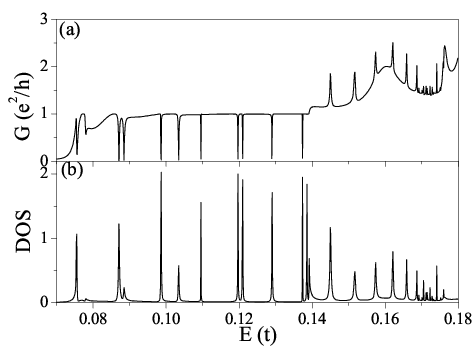}
\caption{ (a) The conductance and (b) DOS versus Fermi energy at $N_x=100$. Other parameters are the same as these in Fig.8(b).}
\end{figure}

As discussed in preceding subsection, the band bending could lead to the zero-conductance resonance. Fig. 10 shows (a) the conductance and (b) DOS calculated by Hubbard model, corresponding to the band structure in Fig. 8(b). One can clearly see that many resonance dips appear in the conductance, that is zero-conductance resonance, corresponding to the peaks of DOS. The result is similar to that calculated by the tight-binding model in Figs. 4 and 5.

\section{Conclusion}

In summary, we have systematically studied the band bending, band gap, and zero-conductance Fano resonance controlled by the edge electric fields, Hubbard interaction, and the intrinsic SOI in a ZSiNRs by the NEGF method. The antisymmetric edge fields could induce a controllable band bending region where the highest valence band and the lowest conduction band coexist. As a consequence, the conductance presents many resonance dips with $G=0$ and resonance peaks with $G=2e^2/h$, which relate to Fano resonance effect. The resonance matches very well with the DOS and the LDOS. Based on the zero-conductance resonances, a perfect spin filter could be achieved. The symmetric edge fields could open an asymmetric band gap at the two valleys controlled by the intrinsic SOI, which could lead to a valley polarized quantum spin Hall state. Furthermore, similar to the edge electric fields, the Hubbard interaction could be regarded as a spin-dependent scalar field acting on the edges. Together with the edge electric fields, the Hubbard interaction could give rise to a phase transition from semiconductor to metal or half-metal.

Compared with previous literature on Fano resonance effect in two-dimensional materials, the model produced by the electric fields in this paper is simpler and easier to be implemented experimentally. In addition, the obtained results can be generalized to other two-dimensional materials such as germanene and stanene. Finally, We hope the results such as the electrically tunable band bending and zero-conductance resonance, will benefit the understanding and application of silicene.

This work was supported by the NSFC (Grants No. 11974153 and No. 11921005), National Key R and D Program of China (Grant No. 2017YFA0303301), the Strategic Priority Research Program of Chinese Academy of Sciences (Grant No. XDB28000000), and the Natural Science Foundation of Shandong Province (Grant No. ZR2017JL007). Email address of Wei-Tao Lu: physlu@163.com.

%\end{CJK*}
\end{document}